\documentclass[12pt]{article}

\usepackage{epsfig}
\usepackage{amsbsy}
\usepackage{amsmath}

\begin{document}

\title{Evaluation of the coincidence probabilities in a generalized Gaussian model of multiple particle production}
\author{A.Bialas and K.Zalewski
\\ M.Smoluchowski Institute of Physics
\\ Jagellonian University, Cracow
\\ and\\ Institute of Nuclear Physics, Cracow}
\maketitle

\begin{abstract}
Coincidence probabilities, which yield Renyi entropies, are investigated in a
generalized Gaussian model, which includes interparticle correlations.
\end{abstract}
\noindent PACS numbers 25.75.Gz, 13.65.+i \\Bose-Einstein correlations,
interaction region determination.

\section{Introduction}

Let us consider an $M$-particle state and assume that its density matrix is

\begin{equation}\label{defden}
  \rho(\textbf{K},\textbf{q}) = e^{-v(\textbf{K}) -\frac{1}{2}\textbf{q}\textbf{L}^2\textbf{q}+ i\textbf{q}\overline{\textbf{X}}(\textbf{K})},
\end{equation}
where $\textbf{K} = \frac{1}{2}(\textbf{p} + \textbf{p'})$, $\textbf{q} =
\textbf{p} - \textbf{p'}$ and $\overline{\textbf{X}}(\textbf{K})$ are
$3M$-dimensional vectors, and $\textbf{L}^2$ is a $3M\times 3M$ dimensional
matrix. Matrix $\textbf{L}^2$ is in general a function of $\textbf{K}$. The
physical interpretation of formula (\ref{defden}) is naturally obtained when
the density matrix is converted into the corresponding Wigner function:

\begin{equation}\label{}
  W(\textbf{K},\textbf{X}) = \frac{\sqrt{2\pi}^{3M}}{Det L}e^{- v(\textbf{K})-\frac{1}{2}(\textbf{X} -
  \overline{\textbf{X}}(\textbf{K}))\textbf{L}^{-2}(\textbf{X} - \overline{\textbf{X}}(\textbf{K}))}
\end{equation}
Note that $Det \textbf{L}$ is an effective volume of the system.

In this paper we derive, assuming (\ref{defden}), an asymptotic formula, valid
when the eigenvalues of matrix $\textbf{L}$ are large, for the coincidence
probabilities

\begin{equation}\label{cldefi}
C(l) = Tr \rho^l =\int\!\! d^lp\; \prod_{j=1}^l
\rho(\textbf{K}_j,\textbf{q}_j),
\end{equation}
which are needed to obtain the Renyi entropies

\begin{equation}\label{}
  H_l = \frac{1}{1-l}\log C(l).
\end{equation}
Here $\textbf{K}_j = \frac{1}{2}(\textbf{p}_j + \textbf{p}_{j+1})$,
$\textbf{q}_j = \textbf{p}_j - \textbf{p}_{j+1}$ and $\textbf{p}_{l+1} \equiv
\textbf{p}_1$.

As seen from formula (\ref{defden}), when the eigenvalues of matrix $L$ are
large, the integral is dominated by the region $q \approx 0$. Therefore, we
propose to expand the exponent in the integrand in powers of the components of
$\textbf{q}$ and to keep only the terms up to second order. Thus, the problem
reduces to the evaluation of a Gaussian integral.

Our final result is an elegant formula generalizing the formulae derived for
the one-dimensional Gaussian model by a number of authors \cite{PRA}-
\cite{ZAL}. For the convenience of the reader, a short derivation of the
formula for the one-dimensional case is given in Section 3. In Section 2 the
integral (\ref{cldefi}) is worked out. In Section 4 this integral is
significantly simplified using the result derived in Section 3. Section 5
contains our conclusions. Two Appendices contain discussions of the matrices
$\textbf{R}^2$ and $S$ introduced in the text.

\section{Evaluation of $C(l)$}

In order to calculate the coincidence probabilities it is convenient to
introduce the notation

\begin{equation}\label{}
  \textbf{K}_j = \overline{\textbf{K}} + \textbf{k}_j,\qquad \overline{\textbf{K}} =
  \frac{1}{l}(\textbf{p}_1+\cdots + \textbf{p}_l).
\end{equation}

Note that

\begin{equation}\label{condkq}
  \sum_{j=1}^l \textbf{k}_j = 0;\qquad \sum_{j=1}^l \textbf{q}_j = 0.
\end{equation}
An immediate implication is that the terms with $\overline{X}$ do not
contribute. Indeed, let us consider the expansions of each term
$q_j\overline{\textbf{X}}(\textbf{K}_j)$ in powers of the components of
$\textbf{q}$. The first terms are
$q_j\overline{\textbf{X}}(\overline{\textbf{K}})$ and their total contribution
to the exponent is

\begin{equation}\label{}
 i \overline{\textbf{X}}(\overline{\textbf{K}})\sum_{j=1}^l\textbf{q}_j = 0.
\end{equation}
The total contribution from the second terms of the expansion must be zero
because of the hermiticity of the density matrix and all the further terms are
negligible being cubic or higher order in the components of $\textbf{q}$.

It is convenient to replace in (\ref{cldefi}) the variables
$\textbf{p}_1,\ldots,\textbf{p}_l$ by the variables
$\overline{\textbf{K}},\textbf{q}_1,\ldots,\textbf{q}_{l-1}$. The Jacobian of
this change of variables equals one. Further, one introduces into the integrand
the factor

\begin{equation}\label{}
  \delta(\sum_{j=1}^l \textbf{q}_j) =
  \int\!\!\frac{dt}{(2\pi)^{3M}}
  e^{-i\textbf{t}\sum_{j=1}^l \textbf{q}_j},
\end{equation}
where $t$ is another $3M$-dimensional vector, and compensates it by the
integration over $\textbf{q}_j$. Thus, (\ref{cldefi}) takes the form

\begin{equation}\label{}
  C(l) = \int\!\!dK\;\int\!\!\frac{dt}{(2\pi)^3M}\int\!\!d^lq
  e^{\sum_{j=1}^l\left(-v(\textbf{K}_j) - \frac{1}{2}\textbf{q}_jL^2\textbf{q}_j -
  i\textbf{t}\textbf{q}_j\right)}.
\end{equation}

The sum of the $v$ terms can be rewritten as follows

\begin{equation}\label{}
  \sum_{j=1}^lv(\textbf{K}_j) =
  lv(\overline{\textbf{K}})+ {\frac{1}{2}\sum_{j=1}^l(\textbf{k}_j\nabla)(\textbf{k}_j
  \nabla) v(\overline{\textbf{K}})},
\end{equation}
where the differential operators $\nabla_\alpha$ act on the components of
$\overline{\textbf{K}}$. Expressing the vectors $\textbf{k}_j$ as linear
combinations of the vectors $\textbf{q}_j$ and performing the summation over
$j$ one finds that

\begin{equation}\label{}
\sum_{j=1}^lv(\textbf{K}_j) =
  lv(\overline{\textbf{K}}) +
 \frac{1}{2}\sum_{j,k}^{l-1}
 \textbf{q}_jS_{jk}
 \textbf{R}^2\textbf{q}_k,
\end{equation}
where $\textbf{R}^2$ is a $3M\times 3M$ matrix with elements

\begin{equation}\label{rkwdef}
  R^2_{\alpha \beta} = \nabla_\alpha
  \nabla_\beta v(\overline{\textbf{K}})
\end{equation}
and $S$ is an $l\times l$ matrix. Matrix $\textbf{R}^2$ for a specific model is
discussed in Appendix 1. The matrix elements $S_{jk}$ are not needed for our
calculation. They are, however, useful for cross-checks. Therefore, they are
calculated and discussed in Appendix 2. Actually, as seen there, there are
potentially useful ambiguities in the definitions of the matrices $S$.

The next step is to replace the integration variables $\textbf{q}_j$ by their
linear combinations $\textbf{Q}_j$, which are the eigenvectors of matrix $S$:
$\textbf{q}_j = \sum_{k=1}^l U_{jk}\textbf{Q}_k$. This transformation can be
chosen orthogonal and then its Jakobian equals one. Using the notation

\begin{equation}\label{}
  U^T S U = \Lambda;\qquad \Lambda_{jk} = \Lambda_j\delta_{jk}
\end{equation}
one finds

\begin{equation}\label{}
  C(l) = \int\!\!d\overline{K}\;
  e^{-lv(\overline{\textbf{K}})}\int\!\!\frac{dt}{(2\pi)^{3M}}\;\prod_{j=1}^l\;\int\!\!dQ\;
 \exp\left[-\frac{1}{2} \textbf{Q}(\textbf{L}^2 + \Lambda_j \textbf{R}^2)\textbf{Q} + i\textbf{t}\sum_{k=1}^l U_{kj}\textbf{Q}\right].
\end{equation}

Further, for each $j$ we introduce an orthogonal transformation $\textbf{Q} =
V^{(j)}\textbf{z}$ which diagonalizes the quadratic form in the exponent

\begin{equation}\label{defmal}
  V^{(j)T}(\textbf{L}^2+\Lambda_j \textbf{R}^2)V^{(j)} = \lambda^{(j)};\qquad \lambda^{(j)}_{\alpha \beta} =
  \lambda^{(j)}_\alpha \delta_{\alpha \beta}.
\end{equation}
This reduces each integral over $Q$ into a product of $3M$ single Gaussian
integrals. Performing them one gets

\begin{equation}\label{clprim}
  C(l) = \sqrt{2\pi}^{3M(l -
  2)}\int\!\!d\overline{\textbf{K}}\;\frac{e^{-lv(\overline{\textbf{K}})}}{\sqrt{\prod_{j=1}^l Det(\textbf{L}^2 + \Lambda_j
  \textbf{R}^2)}}
  \int\!\!dt
  \exp\left[-\frac{l}{2}\sum_{j=1}^l C_j\textbf{t}V^{(j)}\frac{1}{\lambda^{(j)}}V^{(j)T}\textbf{t}\right],
\end{equation}
where
\begin{equation}\label{}
 C_j = \frac{1}{l}\sum_{ik}U_{ij}U_{kj}
\end{equation}
and the identities

\begin{equation}\label{}
  \prod_{\alpha = 1}^{3M}\lambda^{(j)}_\alpha = Det (\textbf{L}^2 + \Lambda_j \textbf{R}^2)
\end{equation}
have been used.

Performing in (\ref{clprim}) the integration over $t$ and extracting $Det
\textbf{L}^2 = (Det \textbf{L})^2$ one gets

\begin{equation}\label{}
 C(l) = \left(\frac{\sqrt{2\pi}}{Det \textbf{L}}\right)^{3M(l -
  1)}l^{-\frac{3M}{2}}\int\!\!d\overline{\textbf{K}}\;\frac{e^{-lv(\overline{\textbf{K}})}}{\sqrt{Det \textbf{A}\prod_{j=1}^l Det(1 +
  \Lambda_j\textbf{L}^{-1}\textbf{R}^2\textbf{L}^{-1})}},
\end{equation}
where

\begin{equation}\label{}
  \textbf{A} = l\sum_{j=1}^l C_j \textbf{L}^{-1}V^{(j)}\frac{1}{\lambda^{(j)T}}V^{(j)T}\textbf{L}^{-1}.
\end{equation}
Using (\ref{defmal}) this can be also rewritten as

\begin{equation}\label{}
  \textbf{A} = l\sum_{j=1}^l \frac{C_j}{1 + \Lambda_j
  \textbf{L}^{-1}\textbf{R}^2\textbf{L}^{-1}}
\end{equation}
which yields

\begin{equation}\label{clwste}
  C(l) = \left(\frac{\sqrt{2\pi}}{Det \textbf{L}}\right)^{3M(l -
  1)}l^{-\frac{3M}{2}}
  \int\!\!d\overline{\textbf{K}}\;
  \frac{e^{-lv(\overline{\textbf{K}})}}
  {\sqrt{Det\left[ \sum_{j=1}^l C_j
  \prod_k^{l(j)}(1 +
  \Lambda_k\textbf{L}^{-1}\textbf{R}^2\textbf{L}^{-1})\right]}}.
\end{equation}
The symbol $\prod_{k=1}^{l(j)}$ denotes the product $\prod_{k=1}^l$ with the
$j$-th factor omitted. This is a usable expression for the coincidence
probabilities $C(l)$\footnote{It requires, however, the determination of the
eigenvalues $\Lambda_k$ and of the coefficients $C_j$. This is a nontrivial
task for large $l$.}. The formula, however, can be significantly simplified. In
order to derive the simplification we will need a result concerning the
one-dimensional Gaussian model. This is derived in the following section.

\section{The one-dimensional Gaussian model}

The model considered in the present paper is a generalization of the more
common Gaussian model, where the $M$-particle density matrix is a product of
$3M$ one dimensional Gaussian density matrices of the form:

\begin{equation}\label{gausim}
  \rho(K,q) = \frac{R}{\sqrt{2\pi}}e^{-\frac{1}{2}R^2K^2 - \frac{1}{2}L^2q^2},
\end{equation}
where $K,q,R,L$ are just real numbers.  Let us calculate the coincidence
probabilities $C_1(l)$ for the one-dimensional model (\ref{gausim}). To this
end we observe that $\rho(K,q)$ is diagonal in the representation of the wave
functions of the harmonic oscillator \cite{BZ1}-\cite{ZAL} and can be written
in the form

\begin{equation}\label{rhodia}
  \rho(K,q) = \sum_{n=0}^\infty \psi_n(p)\lambda_n\psi_n^*(p'),
\end{equation}
where

\begin{eqnarray}\label{}
\lambda_n &=& (1-z)z^n;\\ \psi_n(p) &=& \sqrt{\frac{\alpha}{\sqrt{\pi}2^n n!}}
  \exp\left(-\frac{1}{2}\alpha^2p^2\right)H_n(\alpha p); \qquad \\
  \label{alphaz}
  \alpha &=& \sqrt{RL};\qquad z = \frac{1 -\frac{R}{2L}}{1+\frac{R}{2L}}.
\end{eqnarray}

 Therefore,

\begin{equation}\label{trrhol}
  Tr\rho^l = \sum_{n=0}^\infty \lambda_n^l = (1-z)^l\frac{1}{1 - z^l},
\end{equation}
where a geometric progression has been summed. It is seen from  (\ref{alphaz})
that $|z| < 1$, as it should.

To show that formulae (\ref{gausim}) and (\ref{rhodia}) are equivalent, one may
use the identity\footnote{To prove (\ref{hermit}) it is enough to substitute on
the left hand side twice the definition $H_n(u) = \frac{2^n}{\sqrt{\pi}}
\int_{-\infty}^\infty\!\!dt\;(u+it)^n e^{-t^2}$, perform the summation and a
Gaussian integration.}

\begin{equation}\label{hermit}
  \sum_{n=0}^\infty\frac{z^n}{n!}H_n(x)H_n(y) =
  \frac{1}{\sqrt{1-4z^2}}\exp\left(-4z\frac{z(x^2 + y^2) -xy}{1-4z^2}\right)
\end{equation}

Substituting the formula for $z$ into (\ref{trrhol}) one easily finds

\begin{equation}\label{c1exac}
  C_1(l) =
  \left(\frac{R}{L}\right)^l
  \frac{1}{\left(1 + \frac{R}{2L}\right)^l -
 \left(1 - \frac{R}{2L}\right)^l}.
\end{equation}
This is the formula we need.

\section{Final expression for $C(l)$}

The one-dimensional Gaussian model can be also studied using the methods from
Section 2. There is a number of simplifications, however. Since the dimension
of the matices $3M$ is replaced by one

\begin{equation}\label{}
  \lambda^{(j)} = L^2 + \Lambda_jR^2;\qquad (d=1).
\end{equation}
Since there is no need to diagonalize $L^2 + \Lambda_iR^2$: $V^{(j)} =
V^{(j)T}= 1$. Moreover,

\begin{equation}\label{}
  e^{-lv(\overline{K})} =
  \left(\frac{R}{\sqrt{2\pi}}\right)^le^{-\frac{l}{2}R^2\overline{K}^2};\qquad (d=1).
\end{equation}
Thus

\begin{equation}\label{cgaus1}
  C_1(l) = \frac{R^l}{2\pi\sqrt{\prod_{j=1}^l(L^2 + \Lambda_jR^2)}}
  \int\!\!dK
  e^{-\frac{l}{2}K^2R^2}\int\!\!dt\exp\left[-\frac{l}{2}\sum_{j=1}^l C_j
  \frac{t^2}{L^2+\Lambda_jR^2}\right].
\end{equation}
Note that in spite of all these simplifications the eigenvalues $\Lambda_j$ and
the coefficients $C_j$ are the same as in the general case. Performing the
integrations over $t$ and $K$:

\begin{equation}\label{}
  C_1(l) = \frac{R^{l-1}}{\sqrt{l}\sqrt{A \prod_{j=1}^l(L^2 + \Lambda_jR^2)}},
\end{equation}
where

\begin{equation}\label{}
  A = l\sum_{j=1}^l\frac{C_j}{L^2 + \Lambda_jR^2};\qquad (d=1).
\end{equation}
The formula for $C_1(l)$ can be rewritten as

\begin{equation}\label{c1news}
  C_1(l) = \left(\frac{R}{L}\right)^{l-1}\frac{1}{l}\frac{1}{\sqrt{\sum_{j=1}^l C_j
  \prod_{k=1}^{l(j)}(1 + \Lambda_j\left(\frac{R}{L}\right)^2)}},
\end{equation}
 Let us compare now this result with formula (\ref{c1exac}). The
equivalence of the two formulae implies that

\begin{equation}\label{}
l\sqrt{\sum_{j=1}^l C_j
  \prod_{k=1}^{l(j)}(1 + \Lambda_j\left(\frac{R}{L}\right)^2)}=
  \frac{L}{R}\left[\left(1 + \frac{R}{2L}\right)^l - \left(1 -
  \frac{R}{2L}\right)^l\right],
\end{equation}
or equivalently

\begin{equation}\label{exproo}
\sqrt{\sum_{j=1}^l C_j
  \prod_{k=1}^{l(j)}(1 + \Lambda_j\left(\frac{R^2}{L^2}\right))}=
  1 + \sum_{n=1}^{N(l)} 2^{-2n}\frac{(l-1)!}{(2n+1)!(l-1-2n)!}\left(\frac{R^2}{L^2}\right)^n,
\end{equation}
where $N(l) = E\left[\frac{1}{2}(l-1)\right]$. Note that this identity is valid
whatever is substituted for $\frac{R^2}{L^2}$.

Let us now go back to the general case. The argument of the determinant in
formula (\ref{exproo}) differs from the argument of the square root in formula
(\ref{c1news}) only by the substitution of the matrix
$\textbf{L}^{-1}\textbf{R}^2\textbf{L}^{-1}$ for the number $\frac{R^2}{L^2}$.
Therefore, the same substitution can be made in the identity (\ref{exproo}) and
one obtains

\begin{equation}\label{clfina}
  C(l) = \left(\frac{\sqrt{2\pi}}{Det \textbf{L}}\right)^{3M(l-1)}
  l^{-\frac{3M}{2}} \int\!\!dK\;
  \frac{e^{-lv(K)}}{Det\left[1 + \sum_{n=1}^{N(l)}a_n
  \left(\textbf{L}^{-1}\textbf{R}^2\textbf{L}^{-1}\right)^n\right]},
\end{equation}
where
\begin{equation}\label{}
  a_n = 2^{-2n}\frac{(l-1)!}{(2n+1)!(l-1-2n)!}.
\end{equation}
This is our final formula and the main result of this paper. Let us note that
when the eigenvalues $t_1,\ldots,t_{3M}$ of matrix

\begin{equation}\label{}
  \textbf{V} = \textbf{L}^{-1}\textbf{R}^2\textbf{L}^{-1}
\end{equation}
are known, the determinant in the integrand of formula (\ref{clfina}) can be
evaluated and the we get

\begin{equation}\label{clweig}
C(l) = \left(\frac{\sqrt{2\pi}}{Det \textbf{L}}\right)^{3M(l-1)}
  l^{\frac{3M}{2}} \int\!\!dK\;
  \frac{e^{-lv(K)\sqrt{\prod_{\alpha=1}^{3M}t_\alpha}}}{\prod_{\alpha = 1}^{3M}
  \left[(1+\frac{1}{2}\sqrt{t_\alpha})^l- (1-\frac{1}{2}\sqrt{t_\alpha})^l\right]}.
\end{equation}

\section{Discussion}

We have derived an explicit formula for the coincidence probabilities in a
general, multidimensional Gaussian model. This model can be considered as an
approximation, at large volume of the system, of a model with an arbitrary
momentum distribution. Therefore, our result may be useful for a rather wide
class of physical situations, particularly for analyses of the systems created
in heavy ion collisions. The formula obtained in the present paper is elegant,
transparent and easy to use.

Some comments are in order.
\begin{itemize}
  \item Since we do not assume factorization of the multiparticle density
  matrix, our calculation takes into account possible correlations between
  particles. Correlations show up as the non-diagonal terms in the matrix
  $\textbf{R}^2$. There is no restriction on their character and magnitude.
  \item It should be emphasized that, although the model is considered as an
  asymptotic expansion for large $\textbf{L}$, the terms in eq. (\ref{clfina}) of higher order in
  $\textbf{L}^{-2}$ should, in general, be included in spite of the fact that terms of
  these orders have been already neglected in the exponent of (\ref{defden}).
  The point is that in the presence of correlations some eigenvalues of matrix
  $\textbf{L}^{-1}\textbf{R}^2\textbf{L}^{-1}$ may be of the order of the number of particles $M$.
  Then the expansion in formula (\ref{clfina}) contains terms of the order
  $\left(\frac{M}{L^2}\right)^n$, which for high multiplicities are not
  necessarily small even for a large system.
\end{itemize}

In order to illustrate the last point consider the
matrices\footnote{The model could be made more realistic by
distinguishing between the interparticle correlations and the
correlations between the directions $x,y,z$. This, however,  would
not change the qualitative results we need.}

\begin{equation}\label{}
 R^2_{ij} = \alpha(\delta_{ij} + \beta); \qquad L_{ij} =
L\delta_{ij}\qquad i,j = 1,\ldots,3M.
\end{equation}
For these matrices the eigenvalues of matrix
$\textbf{L}^{-1}\textbf{R}^2\textbf{L}^{-1}$ are

\begin{equation}\label{}
\lambda_1 = \frac{\alpha(1 +3M\beta)}{L^2};\qquad \lambda_2 = \cdots =
 \lambda_{3M} = \frac{\alpha}{L^2}.
\end{equation}
Therefore,

\begin{equation}\label{}
Det\left[1 + \sum_{n=1}^{N(l)}a_n
  \left(\textbf{L}^{-1}\textbf{R}^2\textbf{L}^{-1}\right)^n\right] =
  \left[1 + \sum_{n=1}^{N(l)}a_n \left(\frac{\alpha(1
  +3M\beta)}{L^2}\right)^n\right]
  \left[1 + \sum_{n=1}^{N(l)}a_n \frac{\alpha^n}{L^{2n}}\right]^{3M-1}
\end{equation}
which shows that terms of the kind $\left(\frac{3M\alpha\beta}{L^2}\right)^n$
do occur in the expansion.

\section{Appendix 1: Matrix $V = L^{-1}R^2L^{-1}$ in a model}

Let us consider a model where

\begin{eqnarray}\label{vjexam}
v(\textbf{K}) &=& \sum_{n=1}^M v(K_n),\nonumber\\
v(K) &=& \frac{1}{T}\sqrt{K_T^2 + m^2} + \frac{Y^2}{2A} + \log\left[\tilde{A}E\right],\\
\tilde{A} &=& \pi T(m+T)\sqrt{8\pi A}\;e^{- \frac{m}{T}}\nonumber.
\end{eqnarray}
In this formula $T$ and $A$ are parameters, $m$ is the mass of the particle;
$K$ are the three-dimensional momentum vectors composing the $3M$ dimensional
vector $\overline{\textbf{K}}$, $K_{T}$ is the transverse component of $K$; the
energy $E = \sqrt{m^2 + K^2}$, the rapidity $Y =
\frac{1}{2}\log\frac{E+K_z}{E-K_z}$. This is a commonly used model, where the
particles are uncorrelated, but there are some correlations between the $x,y,z$
components of the momentum vectors. Thus, matrix $\textbf{R}^2$ consists of $M$
diagonal $3\times 3$ blocks which will be denoted $R^2$. The transverse momenta
have a Boltzmann distribution, while in the longitudinal direction there is a
mild cut-off by a Gaussian in rapidity. The last term in the expression for $v$
ensures the correct normalization. Matrix $\textbf{L}$ is assumed to be
diagonal with diagonal elements $L_x = L_y \equiv L_T,L_z$ in every diagonal
block corresponding to the diagonal blocks of $\textbf{R}^2$. Further we use
the labels $\alpha, \beta $ for the components $x,y,z$ and the labels $a,b$ for
the transverse components $x,y$, the transverse mass $m_T = \sqrt{m^2 +
K_T^2}$.

Let us define

\begin{equation}\label{}
V_{\alpha\beta} = \frac{R^2_{\alpha\beta}}{L_\alpha L_\beta}.
\end{equation}
This gives

\begin{eqnarray}\label{mactab}
V_{ab} &=& \Theta K_a K_b + \Phi\delta_{ab}, \\
 \label{mactaz}
 V_{az} &=& \omega K_a, \\
\label{mactzz}
 V_{zz} &=& \Psi,
\end{eqnarray}
where

\begin{eqnarray}\label{}
  L_T^2\Theta &=& -\frac{1}{Tm_T^3} - \frac{2}{E^4} +
  \frac{K_z}{AEm_T^2}\left(\frac{K_z}{Em_T^2}+ \frac{2Y}{m_T^2} +
  \frac{Y}{E^2}\right),\\
  L_T^2\Phi &=& \frac{1}{Tm_T} + \frac{1}{E^2} - \frac{YK_z}{AEm_T^2}, \\
  L_TL_z\omega &=&  -\frac{1}{AE^2}\left(\frac{K_z}{m_T^2} + \frac{Y}{E}\right) - \frac{2K_z}{E^4}, \\
L_z^2 \Psi &=&  \frac{1}{E^2}\left[\frac{A+1}{A} - \frac{K_z}{E}\frac{2AK_z +
EY}{EA}\right].
\end{eqnarray}
As seen from (\ref{mactab})- \ref{mactzz})  matrix $T$ can be written in the
form

\begin{equation}\label{}
  V_{\alpha\beta} = \Phi\delta_{\alpha\beta} + V'_{\alpha\beta}.
\end{equation}
The second row of matrix $V'$ is proportional to the first. Therefore, $Det V'
= 0$. Consequently, one of the eigenvalues of matrix $V$ equals $\Phi$ and the
other two can be found by solving a quadratic equation. The result is

\begin{eqnarray}\label{}
t_1 &=& \Phi,\\
t_2 &=& \frac{1}{2}\left(\Theta K_T^2 + \Phi + \Psi + \sqrt{(\Theta K_T^2+\Phi-\Psi)^2+4\omega^2K_T^2}\right),\\
t_3 &=& \frac{1}{2}\left(\Theta K_T^2 + \Phi + \Psi - \sqrt{(\Theta
K_T^2+\Phi-\Psi)^2+4\omega^2K_T^2}\right).
\end{eqnarray}
Since the eigenvalues are known formula (\ref{clweig}) can be used to calculate
$C(l)$. In particular for $l=3$ the result is

\begin{equation}\label{}
C(3) = \left(\frac{\sqrt{2\pi}}{L_T^2L_z}\right)^{6M}3^{-\frac{3M}{2}}
 \int\!\!dK\;\frac{e^{-3v(K)}}
 {
  \left[
  (1+\frac{\Phi}{12})
  \left(
  1+\frac{\Theta K_T^2 +\Phi + \Psi}{12} +
  \frac{(\Theta K_T^2 +\Phi)\Psi - \omega^2K_T^2}{144}
  \right)
  \right]^M
  }.
\end{equation}

\section{Appendix 2: Evaluation of the matrix S}

Let us represent each vector $\textbf{k}_j,\quad j=1,\ldots,l$ as a linear
combination of the vectors $\textbf{q}_j,\quad j=1,\quad,l-1$

\begin{equation}\label{}
  \textbf{k}_j = \frac{1}{2}(\textbf{p}_j + \textbf{p}_{j+1}) - \frac{1}{l}(\textbf{p}_1 + \cdots + \textbf{p}_l) =
  \sum_{k=1}^{l-1} c_{jk}\textbf{q}_k.
\end{equation}
Because of the second identity (\ref{condkq}) it is not mandatory to include
$q_l$ in the expansion. Equating the coefficients of the vectors
$\textbf{p}_1,\ldots \textbf{p}_l$ on both sides of the second equality we get
the equation system

\begin{equation}\label{}
  \frac{1}{2}(\delta_{jn} + \delta_{j+1,n}) - \frac{1}{l} = c_{jn} -
  c_{j,n-1};\qquad n = 1,\ldots,l,
\end{equation}
where $c_{j0} = c_{j,l} = 0$ and $\delta_{l+1,n}$ stands for $\delta_{1,n}$.
Summing each side of the equations over $n$ from $n=k+1$ to $n=l$ one gets from
the resulting equalities

\begin{equation}\label{}
  c_{jk} = -\frac{k}{l} + 1 + \frac{1}{2}\delta_{jl} - \frac{1}{2}\left[\Theta(j-k)
  + \Theta(j-k+1)\right],
\end{equation}
where $\Theta(n)$ equals one for $n>0$ and zero otherwise.

The coefficients $S_{mn}$ are defined by

\begin{equation}\label{kanaqu}
  k_{j\alpha}k_{j\beta} = \sum_{m,n}^l S_{mn}q_{m\alpha}q_{n\beta}.
\end{equation}
Substituting on the left hand side the expansions of $\textbf{k}_j$ in terms of
the $\textbf{q}_j$ one finds

\begin{equation}\label{}
  S_{mn} = \sum _{j=1}^l c_{jm}c_{jn}.
\end{equation}
Note that this matrix is symmetric and that by construction $S_{ml} = S_{lm} =
0$. Using the explicit formulae to perform the summation

\begin{equation}\label{forsmn}
  S_{mn} = \frac{m(l-n)}{l} - \frac{1}{4}(1 + \delta_{mn}),\quad
  \mbox{for}\quad m \leq n < l.
\end{equation}
For $m > n$ one finds the matrix elements from the symmetry $S_{mn} = S_{nm}$.

For $l =2$ one gets $S = 0$, which greatly simplifies the analysis \cite{BCZ2}.
Note also that because of the second identity (\ref{condkq}) one can add on the
left hand side of relation (\ref{kanaqu}) a term

\begin{equation}\label{}
\sum_{j=0}^l\textbf{q}_j \sum_{k=1}^lc_k\textbf{q}_k,
\end{equation}
where $c_k$ are arbitrary constants. This modifies $S$ without affecting the
products $\textbf{k}_{j\alpha}\textbf{k}_{j\beta}$ and sometimes can be used to
simplify the calculations.

Choosing for instance:  $c_1 = c_2 = -c_3 = -\frac{1}{12}$ for $l=3$ and $c_1 =
c_3 = -c_4 = \frac{1}{3}c_2 = \frac{1}{8}$ for $l=4$ one gets the simple
formulae:

\begin{equation}\label{}
S(3) = \begin{pmatrix}
    \frac{1}{12} & 0 & 0 \\
    0& \frac{1}{12} & 0 \\
    0 & 0 & \frac{1}{12} \
  \end{pmatrix}, \qquad S(4) = \left( \begin{array}{cccc}
    \frac{1}{8} & 0 & -\frac{1}{8} & 0 \\
    0 & \frac{1}{8} & 0 & -\frac{1}{8} \\
    -\frac{1}{8} & 0 & \frac{1}{8} & 0 \\
    0 & -\frac{1}{8} & 0 & \frac{1}{8} \
  \end{array}\right)
\end{equation}
from which the eigenvalues are immediately visible.

\textbf{Acknolwledgements} The contribution of Wies{\l}aw Czy\.{z} at the early stages
of this investigation is highly appreciated. This work has been partly
supported by the Polish Ministry of Education and Science grant 1P03B 045
29(2005-2008).

\end{document}